\documentclass[preprint,12pt]{elsarticle}

\usepackage{bm}

\usepackage{amssymb}

\journal{Physics Letters B}

\begin{document}

\begin{frontmatter}



\title{Shrunk halo and quenched shell gap at $N=16$ in $^{22}$C:
       Inversion of $sd$ states and deformation effects}


\author[ITP,UCAS]{Xiang-Xiang Sun}
\author[CAEP]{Jie Zhao}
\author[ITP,UCAS,HIRFL,HNNU]{Shan-Gui Zhou\corref{cor1}}
  \ead{sgzhou@itp.ac.cn}
  \cortext[cor1]{Corresponding author.}

\address[ITP]{CAS Key Laboratory of Theoretical Physics,
              Institute of Theoretical Physics,
              Chinese Academy of Sciences, Beijing 100190, China}
\address[UCAS]{School of Physical Sciences,
               University of Chinese Academy of Sciences,
               Beijing 100049, China}
\address[CAEP]{Microsystem and Terahertz Research Center,
               China Academy of Engineering Physics,
               Chengdu 610200, Sichuan, China}
\address[HIRFL]{Center of Theoretical Nuclear Physics, National Laboratory
                of Heavy Ion Accelerator, Lanzhou, 730000, China}
\address[HNNU]{Synergetic Innovation Center for Quantum Effects and Application,
               Hunan Normal University, Changsha, 410081, China}

\begin{abstract}
We explore the interplays among 
the formation of a halo, deformation effects,
the inversion of $sd$ states,
the shell evolution, and
changes of nuclear magicities in $^{22}$C
by using a deformed relativistic Hartree-Bogoliubov model in continuum.
It is revealed that there is an inversion between the two spherical orbitals
$2s_{1/2}$ and $1d_{5/2}$ in $^{22}$C compared with the conventional single particle
shell structure in stable nuclei.
This inversion, together with deformation effects,
results in a shrunk halo and a quenched shell gap at $N=16$.
It is predicted that the core of $^{22}$C is oblate but the halo is prolate.
Therefore several exotic nuclear phenomena, including the halo, the shape decoupling
effects, the inversion of $sd$ states, and the evolution of shell structure which
results in (dis)appearance of magic numbers, coexist in one single nucleus
$^{22}\mathrm{C}$.
\end{abstract}


\end{frontmatter}


\section{Introduction}
\label{sec:intro}

The study of exotic nuclear structure is at the forefront of research in
modern nuclear physics \cite{Zhou2017_PoS-INPC2016-373}.
Among many others, the most striking exotic nuclear phenomenon is
the nuclear halo which was first observed in $^{11}$Li
\cite{Tanihata1985_PRL55-2676}. 
Halo nuclei are weakly bound and well associated with pairing correlations
and the contribution of the continuum above the threshold of particle emission
\cite{Hansen1987_EPL4-409,
Dobaczewski1996_PRC53-2809,
Meng1996_PRL77-3963,Meng1998_PRL80-460,Meng1998_NPA635-3,Jensen2004_RMP76-215,
Tanihata2013_PPNP68-215,
Riisager2013_PST152-014001,
Zhang2014_PLB730-30,
Hu2018_in-prep}.
The formation of a nuclear halo is closely connected with the evolution
of the shell structure and changes of nuclear magicities around drip-lines
\cite{Dobaczewski1994_PRL72-981,Meng1998_PLB419-1,Long2010_PRC81-031302R,
Otsuka2018_arXiv1805.06501}.

Most known nuclei are deformed with shapes deviating from a sphere.
When the deformation is involved in, even more exotic phenomena are expected
\cite{Misu1997_NPA614-44}.
The shape decoupling phenomenon, i.e., the core and the halo having different shapes,
has been predicted in deformed nuclei close to the neutron drip-line
\cite{Zhou2010_PRC82-011301R,Li2012_PRC85-024312}.
For example, in $^{42,44}$Mg, 
the core and the halo are predicted to be prolate and oblate, respectively.
Such predictions were made by using a deformed relativistic Hartree-Bogoliubov
model in continuum (DRHBc model)
\cite{Zhou2010_PRC82-011301R,Li2012_PRC85-024312,Li2012_CPL29-042101}
which describes self-consistently the large spatial extension,
the contribution of the continuum due to pairing correlations, and
deformation effects in deformed nuclei with halos.
Later similar shape decoupling effects were also revealed by using
nonrelativistic Skyrme Hartree-Fock-Bogoliubov models
for axially deformed nuclei in coordinate space
\cite{Pei2013_PRC87-051302R,Pei2014_PRC90-051304R,
Xiong2016_ChinPhysC40-024101,Wang2017_PRC96-031301R} or
in a Gaussian basis \cite{Nakada2008_NPA808-47,Nakada2018_PRC98-011301R}.

As the heaviest Borromean nucleus with a halo observed so far, $^{22}$C
is of particular interest because of not only possible new magicities
but also uncertainties and puzzles in
the separation energy, the matter radius, and the halo configuration.
If the $Z=6$ magic number evidenced in neutron-rich C isotopes
\cite{Tran2018_NatureCommu9-1594}
persists in it and the shell gap at $N=16$ is large enough,
$^{22}$C could be a new doubly magic nucleus \cite{Sorlin2008_PPNP61-602}.
The empirical value of the two-neutron separation energy
$S_{2n}$ is 420(940) keV in AME2003 \cite{Audi2003_NPA729-337}
and 110(60) keV in AME2012
\cite{Audi2012_ChinPhysC36-1157,Audi2012_ChinPhysC36-1287,Wang2012_ChinPhysC36-1603}.
In 2012, $S_{2n}$ was determined to be $-0.14 \pm 0.46$ MeV from direct
mass measurements \cite{Gaudefroy2012_PRL109-202503}.
According to the recent AME2016,
$S_{2n} = 35(20)$ keV
\cite{Audi2017_ChinPhysC41-030001,Huang2017_ChinPhysC41-030002,Wang2017_ChinPhysC41-030003}.
The matter radius of $^{22}$C deduced from interaction
cross sections measured in two experiments differ very much:
$r_m = 5.4 \pm 0.9 $ fm in 2010 \cite{Tanaka2010_PRL104-062701} and
$r_m = 3.44\pm 0.08$ fm in 2016 \cite{Togano2016_PLB761-412}.
Recently, the determination of $^{22}$C radius with interaction cross sections
was re-examined by using the Glauber model and $r_m = 3.38\pm 0.10$ fm was extracted
\cite{Nagahisa2018_PRC97-054614}.
In almost all investigations on $^{22}$C
\cite{Horiuchi2006_PRC74-034311,
Abu-Ibrahim2008_PRC77-034607,Abu-Ibrahim2010_PRC81-019901E,Abu-Ibrahim2009_PRC80-029903E,
Tanaka2010_PRL104-062701,
Coraggio2010_PRC81-064303,
Fortune2012_PRC85-027303,
Ershov2012_PRC86-034331,Shulgina2018_PRC97-064307,
Kobayashi2012_PRC86-054604,
Frederico2012_PPNP67-939,Hagen2013_EPJA49-118,
Lu2013_PRC87-034311,
Ogata2013_PRC88-024616,Acharya2013_PLB723-196,
Kucuk2014_PRC89-034607,
Hoffman2014_PRC89-061305R,
Inakura2014_PRC89-064316,
Togano2016_PLB761-412,
Ji2016_IJMPE25-1641003,
Pinilla2016_PRC94-024620,
Souza2016_PRC94-064002,
Suzuki2016_PLB753-199,
Souza2016_PLB757-368,
Nagahisa2018_PRC97-054614},
the two valence neutrons are
assumed to occupy mostly the second $s$ orbital $2s_{1/2}$.
There are strong interplays among $S_{2n}$, $r_m$, and the halo configuration,
see, e.g., Ref.~\cite{Hammer2017_JPG44-103002} for a recent review.
An apparent puzzle arises from these interplays: if the two valence neutrons
occupy $2s_{1/2}$ and
$S_{2n}$ is very small, say, from several tens keV to several hundreds keV,
the radius of $^{22}$C should be much larger than the recent experimental value.

In this work, we study $^{22}$C with the DRHBc model.
It is shown that the $2s_{1/2}$ orbital is a bit lower than the $1d_{5/2}$ orbital
when the spherical symmetry is imposed, i.e.,
these two states are inverted compared with the conventional
shell structure in stable nuclei.
The near degeneracy of $2s_{1/2}$ and $1d_{5/2}$ would lead to a large
shell gap at $N=16$.
However, the ground state of $^{22}$C is deformed.
The inversion of $(2s_{1/2},1d_{5/2})$, together with deformation effects,
results in a shrinkage in the halo and a quenched shell gap
at $N=16$ in $^{22}$C,
thus resolving the puzzles concerning the radius and halo configuration
in this exotic nucleus.
Furthermore, we predict that the core of $^{22}$C is oblate but the halo is prolate,
adding one more candidate of deformed halo nuclei with shape decoupling effects.

\section{The DRHBc model}
\label{sec:model}

The details of the DRHBc model with nonlinear meson-nucleon couplings can be
found in Refs.~\cite{Zhou2010_PRC82-011301R,Li2012_PRC85-024312,Li2012_CPL29-042101}.
The DRHBc model with density-dependent couplings has been developed
by Chen et al. \cite{Chen2012_PRC85-067301}.
Here we only present briefly the formalism for the convenience of the following discussions.
In the DRHBc model, the RHB equation for nucleons \cite{Kucharek1991_ZPA339-23}
\begin{equation}
 \left(
  \begin{array}{cc}
   h_D - \lambda &    \Delta     \\
   -\Delta^*    & -h^*_D+\lambda \\
  \end{array}
 \right)
 \left( { U_{k} \atop V_{k} } \right)
 =
 E_{k}
  \left( { U_k \atop V_{k} }  \right),
 \label{eq:RHB0}
\end{equation}
is solved in a Woods-Saxon (WS) basis \cite{Zhou2003_PRC68-034323}
which can describe the large spatial extension of halo nuclei.
In Eq.~(\ref{eq:RHB0}),
$h_D$ is the Dirac Hamiltonian,
$\lambda$ is the chemical potential,
and
$E_{k}$ and $\left( U_k, \ V_{k} \right)^\mathrm{T}$ are the quasiparticle energy and
wave function.
The pairing potential reads,
\begin{equation}
 \Delta(\bm{r}_1,\bm{r}_2)= 
 V^{pp}(\bm{r}_1,\bm{r}_2)\kappa(\bm{r}_1,\bm{r}_2),
\label{eq:pairing-potential}
\end{equation}
with a density dependent force of zero-range 
\begin{equation}
 V^{pp}(\bm{r}_1,\bm{r}_2)=  V_0\,\delta( \bm{r}_1 - \bm{r}_2 )
   \left(1-\frac{\rho(\bm{r}_1)}{\rho_\mathrm{sat}}\right)
   \frac{1}{2}(1-P^\sigma),
 \label{eq:pairing_force}
\end{equation}
and the pairing tensor $\kappa(\bm{r}_1,\bm{r}_2)$ \cite{Ring1980,Blaizot1985_QTFS}.
In the Dirac Hamiltonian
\cite{Serot1986_ANP16-1,Reinhard1989_RPP52-439,Ring1996_PPNP37-193,
Vretenar2005_PR409-101,Meng2006_PPNP57-470,Niksic2011_PPNP66-519,
Liang2015_PR570-1,Meng2015_JPG42-093101,Meng2016_RDFNS}
\begin{equation}
 h_D =
  \bm{\alpha} \cdot \bm{p} + V(\bm{r}) + \beta (M + S(\bm{r})),
  \label{eq:Dirac0}
\end{equation}
the scalar and vector potentials are expanded in terms of the Legendre polynomials,
\begin{equation}
 f(\bm{r})   = \sum_\lambda f_\lambda({r}) P_\lambda(\cos\theta),\
 \lambda = 0,2,4,\cdots ,
 \label{eq:expansion}
\end{equation}
so are various densities in the DRHBc model.
Note that for triaxially deformed nuclei, the expansion of potentials
and densities should be made in terms of spherical harmonics \cite{Xia2018_in-prep}.

Our calculations are carried out with the covariant density functional PK1
\cite{Long2004_PRC69-034319}.
Since a zero-range interaction (\ref{eq:pairing_force}) is used
in the $pp$ channel, the pairing strength $V_0$ is connected with
a truncation in the quasiparticle space.
The Borromean feature of $^{22}$C is used to fix the pairing
parameters as:
$\rho_\mathrm{sat} =$ 0.152~fm$^{-3}$,
$V_0 = 355$~MeV$\cdot$fm$^3$, and a cut-off energy
$E^\mathrm{q.p.}_\mathrm{cut} = 60$~MeV in the quasi-particle space.
These parameters result in $S_n = -28$ keV for $^{21}$C and
$S_{2n} = 0.43$ MeV for $^{22}$C.

\section{Results and discussions}
\label{sec:results}

\begin{figure}[h]
\begin{center}
\includegraphics[width=0.65\textwidth]{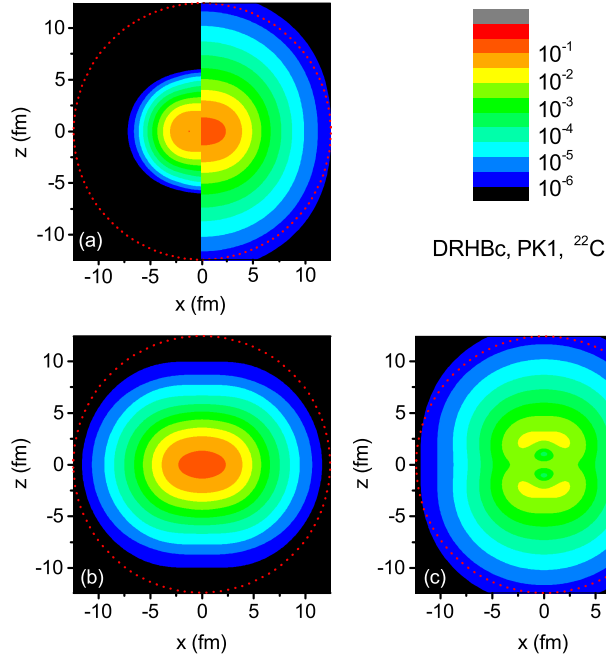}
\end{center}
\caption{(Color online)
Density profiles of $^{22}$C with the $z$ axis as the symmetry axis.
(a) The proton ($x<0$) and neutron ($x>0$) density profiles,
(b) the density profile of the neutron core, and
(c) the density profile of the neutron halo.
In each plot, a dotted circle is drawn to guide the eye.}
\label{fig:den_C22}
\end{figure}

In Fig.~\ref{fig:den_C22}, we display the density profiles of $^{22}$C.
The density distribution of the protons and neutrons are shown in the
left and right parts of Fig.~\ref{fig:den_C22}(a), respectively.
It is clearly seen that the neutrons extend spatially much farther
than the protons, hinting a neutron halo in $^{22}$C.
The calculated matter radius $r_m = 3.25$ fm is significantly smaller
than the experimental value $5.4 \pm 0.9 $ fm given in 2010
\cite{Tanaka2010_PRL104-062701}
but close to the value $3.44\pm 0.08$ fm obtained in 2016 \cite{Togano2016_PLB761-412}
and $3.38\pm 0.10$ fm extracted recently \cite{Nagahisa2018_PRC97-054614}.

It should be mentioned that the empirical radius formula
$r_m = 1.2A^{1/3}$ fm gives $3.36$ fm
for $A=22$ isobars \cite{Kemper2010_Physics3-13}.
This fact indicates that the halo in $^{22}$C is not so pronounced
if we adopt $r_m$ values from Refs.~\cite{Togano2016_PLB761-412,
Nagahisa2018_PRC97-054614} or from our calculations.
Having in mind that the two-neutron separation energy
$S_{2n}$ is quite small ($\le 0.5$ MeV) as we have mentioned,
such ``small'' $r_m$ values are quite puzzling if one accepts
the assumption that the valence neutrons in $^{22}$C occupy mostly
the $2s_{1/2}$ state.
Next we address this issue by examining the halo configuration.

\begin{figure}[htb]
\begin{center}
\includegraphics[width=0.58\textwidth]{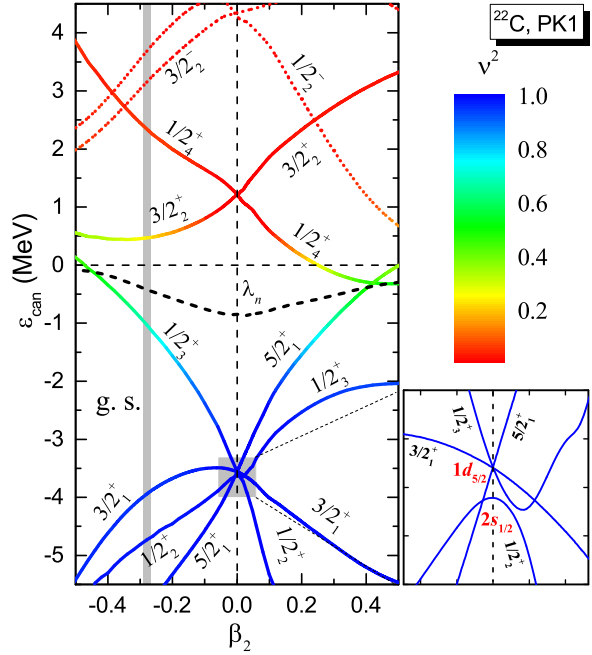}
\end{center}
\caption{(Color online)
Single neutron orbitals around the Fermi level of $^{22}$C
in the canonical basis obtained from constraint calculations.
We label each level with $\Omega^\pi_i$ where $\pi$ is the parity,
$\Omega$ is the projection of angular momentum on the symmetry axis, and
$i$ is used to order the level in the $\Omega^\pi$-block.
The Fermi level ($\lambda_n$) is displayed by the black dashed line.
The occupation probability $\nu^2$ of each orbital is represented with different colors.
The grey vertical line at $\beta_2 = -0.27$ corresponds to the ground state (g. s.) of $^{22}$C.
The shaded region with
$-0.07 \le \beta_2 \le 0.07$ and $-3.8$ MeV $ \le \varepsilon_{\mathrm{can}} \le -3.4 $ MeV
is enlarged and shown on the right side.}
\label{fig:C22-cst-SPL}
\end{figure}

\begin{figure}[h]
\begin{center}
\includegraphics[width=0.65\textwidth]{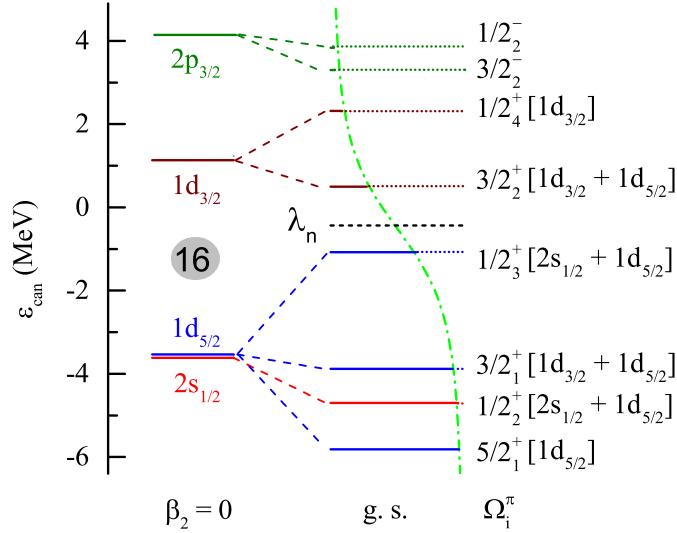}
\end{center}
\caption{(Color online)
Single neutron orbitals around the Fermi level ($\lambda_n$) of $^{22}$C
in the canonical basis in the spherical limit and at the ground state (g. s.).
For the case of the ground state, the length of the solid line is proportional to
the occupation probability of each level calculated from the DRHBc model.
The dash-dotted line corresponds to the occupation probability calculated
from the BCS formula with an average pairing gap.
Quantum numbers $\Omega^\pi_i$ and the main Woods-Saxon components are
given for orbitals in the $sd$ shell.}
\label{fig:C22-SPL}
\end{figure}

The augmented Lagrangian method \cite{Staszczak2010_EPJA46-85} was
implemented in the DRHBc model and deformation constraint calculations
are carried out for $^{22}$C.
In Fig.~\ref{fig:C22-cst-SPL},
we show single neutron levels around the Fermi surface in the canonical basis.
The ground state of $^{22}$C locates at $\beta_2 = -0.27$, as
indicated by the grey vertical line.
There are several orbitals close to the Fermi level and the particle emission threshold:
$1/2^+_3$ is weakly bound and $3/2^+_2$ and $1/2^+_4$ are in the continuum.
These states contribute mostly to the halo and its deformation in $^{22}$C as we will show later.
From Fig.~\ref{fig:C22-cst-SPL} one can find that $1/2^+_3$
becomes more deeply bound
with $\beta_2$ increasing from the ground state
and joins $1d_{5/2}$ 
with $\varepsilon_{\mathrm{can}} \sim -3.6$ MeV at $\beta_2 = 0$.
On the other hand, from the ground state to the spherical limit,
$3/2^+_2$ and $1/2^+_4$ get closer in energy and finally merge as $1d_{3/2}$
which is around 1 MeV above the threshold.
The single neutron levels in the canonical basis in the spherical limit
and at the ground state are also shown in Fig.~\ref{fig:C22-SPL}.

It is interesting to see in Figs.~\ref{fig:C22-cst-SPL} and \ref{fig:C22-SPL} that,
in the spherical limit, 
the $2s_{1/2}$ state is lower than $1d_{5/2}$, i.e.,
these two states are inverted compared with the conventional
shell structure in stable nuclei.
This inversion, together with the large spin-orbit splitting between
the two $d$ states, results in a noticeable shell gap at $N=16$
when $^{22}$C is constrained to be spherical.
The inversion of $(2s_{1/2}, 1d_{5/2})$ has been predicted
in $A/Z \sim 3$ nuclei \cite{Ozawa2000_PRL84-5493}
and the appearance of the $N=14$ and $N=16$ shell closures
is closely related to the competition of $2s_{1/2}$ and $1d_{5/2}$
\cite{Ozawa2000_PRL84-5493,Otsuka2001_PRL87-082502,Cortina-Gil2004_PRL93-062501,
Brown2005_PRC72-057301,Becheva2006_PRL96-012501,Sorlin2008_PPNP61-602,
Kanungo2009_PRL102-152501,Hoffman2009_PLB672-17,
Kanungo2013_PST152-014002,Otsuka2018_arXiv1805.06501}.
In Ref.~\cite{Inakura2014_PRC89-064316}, it is shown that by decreasing
the parameter $t_0$ in the Skyrme interaction SIII, the $2s_{1/2}$ orbital
approaches $1d_{5/2}$ and finally can be lower than the latter in $^{22}$C.

It has been well accepted that the inversion of $(2s_{1/2}, 1d_{5/2})$
results in the formation of the halo in $^{11}$Li
where the $2s_{1/2}$ orbital is close to $1p_{1/2}$
\cite{Sagawa1992_PLB286-7,Meng1996_PRL77-3963,Borge1997_PRC55-R8,Morrissey1997_NPA627-222}.
This inversion, however, plays an opposite role in $^{22}$C:
It hinders the halo formation when we stick to the spherical limit
because the valence neutrons occupy a $d$-wave orbital.
However, there are strong quadrupole correlations which drive $^{22}$C
to be well deformed with $\beta_2 = -0.27$.
On the one hand, the deformation effects mix the $sd$ orbitals,
increase the neutron level densities around the Fermi surface, and
destroy the $N=16$ shell closure as is seen in Figs.~\ref{fig:C22-cst-SPL} and \ref{fig:C22-SPL}.
On the other hand, the mixture of the $sd$ orbitals results in
non-negligible $2s_{1/2}$ components in $1/2^+_3$ and $1/2^+_4$ which are
either weakly bound or in the continuum.
The total amplitude of the $2s_{1/2}$ component is about
25\% in these two $1/2^+$ orbitals.
Having in mind the degeneracy two,
this means that about half of the valence neutrons is of the $2s_{1/2}$ nature.
Therefore the neutron halo in $^{22}$C is shrunk compared with
what it would be if the halo configuration is dominated by $(2s_{1/2})^2$.

\begin{figure}[h]
\begin{center}
\includegraphics[width=0.5\textwidth]{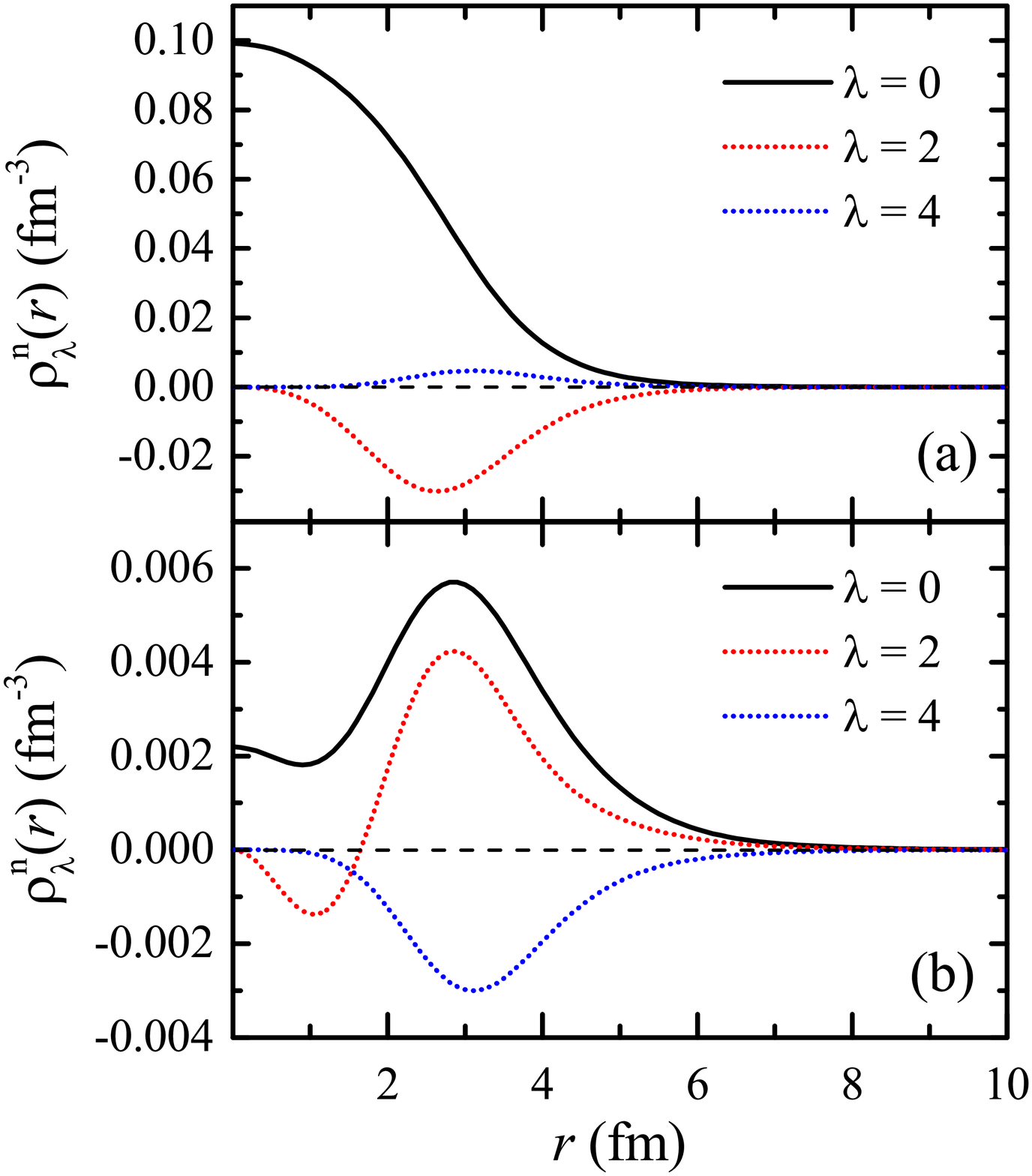}
\end{center}
\caption{Decomposition of the neutron density into
spherical ($\lambda=0$), quadrupole ($\lambda=2$), and hexadecapole
($\lambda= 4$) components for (a) the core and (b) the halo of $^{22}$C.}
\label{fig:den_lambda}
\end{figure}

In Figs.~\ref{fig:C22-cst-SPL} and \ref{fig:C22-SPL}, one can see
that there is a large gap 
between $3/2^+_1$ and $1/2^+_3$.
The orbital $3/2^+_1$ and those below it are deeply bound and contribute to the ``core''.
The orbital $1/2^+_3$ and those above it,
the sum of occupation probabilities of which being 1.03,
are weakly bound or in the continuum and form the ``halo''.
In such a way we can decompose the neutron density into two parts.
The density profiles of the neutron core and halo are presented
in Figs.~\ref{fig:den_C22}(b) and (c), respectively.
It is clearly seen that the core of $^{22}$C is oblate and the
halo is prolate.
This provides one more example of deformed nuclei with a shape decoupling
besides $^{42}$Mg and $^{44}$Mg, both with a prolate core but an oblate halo
\cite{Zhou2010_PRC82-011301R,Li2012_PRC85-024312}.

In Fig.~\ref{fig:den_lambda} the densities of the core and the halo of
$^{22}$C are decomposed into spherical ($\lambda=0$), quadrupole ($\lambda=2$),
and hexadecapole ($\lambda= 4$) components [cf. Eq.~(\ref{eq:expansion})].
In Fig.~\ref{fig:den_lambda}(a), it can be found that the quadrupole
component of the core is always negative, which corresponds to
the oblate shape of $^{22}$C.
However, as seen in Fig.~\ref{fig:den_lambda}(b),
although it is negative when $r<1.6$ fm,
the quadrupole component for the halo is mostly positive,
which is consistent with what we have seen in Fig.~\ref{fig:den_C22}(c),
i.e., the halo of $^{22}$C has a prolate deformation.
From the slope of $\varepsilon_\mathrm{can}$ as a function of $\beta_2$
around the ground state, it can be deduced that the wave function of
the state $1/2^+_{3}$ is prolate and that of $3/2^+_2$ is oblate.
Since it is dominated by $1/2^+_{3}$, the halo in $^{22}$C is prolate.

\section{Conclusions}
\label{sec:summary}

In summary, to resolve the puzzles concerning the radius and configuration
of valence neutrons in $^{22}$C, the ground state properties of $^{22}$C
are studied by using a deformed relativistic Hartree-Bogoliubov model in continuum
with the covariant density functional PK1.
$^{22}$C is predicted to be well deformed with an oblate shape.
The neutrons extend spatially much farther than the protons.
The calculated matter radius $r_m = 3.25$ fm is fairly close to the
two recent experimental values $3.44\pm 0.08$ fm \cite{Togano2016_PLB761-412}
and
$3.38\pm 0.10$ fm \cite{Nagahisa2018_PRC97-054614}
but
much smaller than the experimental value $5.4 \pm 0.9 $ fm
\cite{Tanaka2010_PRL104-062701}.
Deformation constraint calculations reveal that in the spherical limit
the two orbitals $2s_{1/2}$ and $1d_{5/2}$ are inverted in $^{22}$C
compared with the conventional single particle level scheme in stable nuclei.
This inversion hinders the halo formation if $^{22}$C is constrained
to be spherical.
However, strong quadrupole correlations mix the $sd$ orbitals.
This mixture results in sizable $2s_{1/2}$ components in valence neutron
orbitals which are either weakly bound or in the continuum and
leads to a shrunk halo in $^{22}$C.
The deformation effects also increase the neutron level densities
around the Fermi surface and destroy the $N=16$ shell closure.
The neutron density is decomposed into the core and halo.
It is found that the core of $^{22}$C is oblate but the halo is prolate.
Thus this nucleus becomes a new candidate of deformed halo nuclei with
shape decoupling effects.
Finally we mention that the present study was based on the effective
interaction PK1 which is of meson-exchange and it will be interesting
to make similar investigations with point-coupling interactions,
e.g., PC-PK1 \cite{Zhao2010_PRC82-054319}.

\section*{Acknowledgements}

Helpful discussions with S. N. Ershov are gratefully acknowledged.
This work has been supported by
the National Key R\&D Program of China (2018YFA0404402),
the NSF of China (11525524, 11621131001, 11647601, 11747601, and 11711540016),
the CAS Key Research Program (QYZDB-SSWSYS013 and XDPB09),
and
the IAEA CRP ``F41033''.
The computation of this work was supported by
the HPC Cluster of KLTP/ITP-CAS and
the Supercomputing Center, CNIC of CAS.










\end{document}